\documentstyle[12pt]{article}
\begin{document}
\def\om{\omega}
\def\omt{\tilde{\omega}}
\def\ti{\tilde}
\def\o{\Omega}
\def\t{T^*M}
\def\vt{\tilde{v}}
\def\ot{\tilde{\Omega}}
\def\otwo{\omt \wedge \om}
\def\owot{\om \wedge \omt}
\def\w{\wedge}
\def\mt{\tilde{M}}

\def\om{\omega}
\def\omt{\tilde{\omega}}
\def\ss{\subset}

\def\om{\omega}
\def\omt{\tilde{\omega}}
\def\ti{\tilde}
\def\o{\Omega}
\def\t{T^*M}
\def\vt{\tilde{v}}
\def\ot{\tilde{\Omega}}
\def\otwo{\omt \wedge \om}
\def\owot{\om \wedge \omt}
\def\w{\wedge}
\def\mt{\tilde{M}}

\def\om{\omega}
\def\omt{\tilde{\omega}}
\def\ss{\subset}
\def\tpm{T_{P} ^* M}
\def\al{\alpha}
\def\alt{\tilde{\alpha}}
\def\la{\langle}
\def\ra{\rangle}
\def\inop{{\int}^{P}_{P_{0}}{\om}}
\def\th{\theta}
\def\tht{\tilde{\theta}}
\def\inox{{\int}^{X}{\om}}
\def\inotx{{\int}^{X}{\omt}}
\def\st{\tilde{S}}
\def\ls{\lambda_{\sigma}}
\def\p{{\bf{p}}}
\def\pb{{\p}_{b}(t,u)}
\def\pbm{{\p}_{b}}
\def\d{\partial}
\def\d+{\partial_+}
\def\d-{\partial_-}
\def\pat{\partial_{\tau}}
\def\pas{\partial_{\sigma}}
\def\dpm{\partial_{\pm}}
\def\l2{\Lambda^2}
\def\be{\begin{equation}}
\def\ee{\end{equation}}
\def\bea{\begin{eqnarray}}
\def\eea{\end{eqnarray}}
\def\ej{{\bf E}}
\def\ed{{\bf E}^\perp}
\def\si{\sigma}
\def\cg{{\cal G}}
\def\cgt{\ti{\cal G}}
\def\cd{{\cal D}}
\def\ce{{\cal E}}
\def\bz{\bar{z}}
\def\e{\varepsilon}
\def\b{\beta}
\begin{titlepage}
\begin{flushright}
{}~
CERN-TH/95-330\\
hep-th/9512040
\end{flushright}

\vspace{1cm}
\begin{center}
{\Large \bf  Poisson-Lie T-duality}\\
{\Large \bf and Loop Groups of Drinfeld Doubles}\\[20pt]
{\small
{\bf C. Klim\v{c}\'{\i}k}\\
Theory Division CERN \\
CH-1211 Geneva 23, Switzerland \\[10pt]
{\bf P. \v Severa }\\
Department of Theoretical Physics, Charles University, \\
V Hole\v sovi\v ck\'ach 2, CZ-18000 Praha,
Czech Republic\\[20pt] }
\begin{abstract}

A duality invariant first order action is constructed
on the loop group of a Drinfeld double. It  gives at the
same time the description of both of the pair of  $\sigma$-models
related
by Poisson-Lie
T-duality. Remarkably, the action contains a WZW-term on the
Drinfeld
double  not only for conformally invariant $\si$-models.
The resulting actions of the models from the dual pair differ
just by a total
derivative corresponding to an ambiguity in  specifying a
two-form whose
exterior derivative is the WZW three-form. This total
derivative is nothing
but the Semenov-Tian-Shansky symplectic form on the
 Drinfeld double and
it gives directly a generating function of the canonical
transformation
relating the $\si$-models from the dual pair.
\end{abstract}
\end{center}
\vskip 0.5cm
\noindent CERN-TH/95-330\\
December 1995

\end{titlepage}

\section{Introduction}

T-duality is a discrete symmetry of string theory whose physical
interpretation  leads to a remarkable equivalence of seemingly
very
different string backgrounds. Using the $\si$-model language,
a natural
questions arise when a $\si$-model admits a dual one or when two
given  $\si$-models form a dual pair. Of course, truly relevant
setting
for asking these questions is the quantum (conformal) field theory
but already at the classical level a nontrivial preselection
of the  $\si$-models admitting duals  can be made. In fact,  a natural
 criterion
of the classical equivalence of $\si$-models  is an existence of a
 canonical
transformation between the phase spaces of the theories
  preserving the
hamiltonians \cite{AAL1,KS1,OA2}. Having such a pair of the
 $\si$-models
it is then natural to ask whether it can be promoted to a pair of
conformal $\si$-models.

Much activity was devoted in the past few years to the study of the
so-called Abelian T-duality \cite{Busch}--\cite{AAL2}. In this case,
all involved $\si$-models posses an Abelian isometry of their targets,
which plays crucial role in the prescription how to perform duality
transformation leading from one model to its dual. In particular,
the Abelian isometry makes the task of finding the generating function
of the canonical transformation between the models of a dual pair
quite easy \cite{AAL1,KS1,OA2}. Starting from the work
\cite{OQ} (see also \cite{FJ,FT}), there is a growing interest \cite{GR}
in a  non-Abelian generalization of the Abelian T-duality, which
is essentially a duality between $\si$-models defined on targets
being a  Lie group and  its Lie (co)algebra respectively. The model
on the Lie group target does possess the non-Abelian isometry
(the right
multiplication of the target by the elements of the group) and
the duality transformation is realized by the
standard procedure of the gauging the isometry and imposing a
zero-curvature condition on the gauge field by adding the
corresponding Lagrange
multipliers. Integrating out the gauge fields the
dual model is obtained where the Lie (co)algebra coordinates are just
the Lagrange multipliers. This procedure by no means insures an
 isometry
of the dual target, in fact, the dual is generically {\it not} isometric
and it was not known how to perform a `duality' transformation back to
 the original model so the very notion of duality was questionable.

Recently, the present authors have shown that the $\si$-models
on  the Lie group and its (co)algebra are  indeed  {\it dual} to each
other from the point of view of the so-called
Poisson-Lie T-duality \cite{KS2}. The crucial ingredience of our
approach was the abandoning of the requirement of the isometry of the
target space as {\it the} condition of dualizability. Actually we have
shown that the non-Abelian duality exists between $\si$-models
 defined
on the targets of arbitrary two Lie groups whose Lie algebras satisfy
the Lie bialgebra condition \cite{D}. A particular example of such
a pair of groups is just any Lie group and its coalgebra (viewed as
the commutative additive group) mentioned above. The duality
transformation simply exchanges  the roles of the two groups and,
in general, none of the two dual models possesses an isometry.
We have also shown that the Poisson-Lie duality
is the canonical transformation\footnote{For the standard non-Abelian
duality between the $SU(2)$-group and its coalgebra this was shown
already in \cite{CZ} and between  the arbitrary simple group and its
coalgebra in \cite{OA1} and partially in \cite{L}. All these cases are
 special
cases of our construction in \cite{KS2}.} thus the classical criterion
of the equivalence of the $\si$-models is indeed fulfilled.

The idea of T-duality without isometry is becoming quite popular right
now \cite{OA2,EG} though concrete examples, except those of the
 Poisson-Lie
T-duality, are rare\footnote{Very recently a new approach
to T-duality was advanced in \cite{EG} where a simple example of
T-duality
without isometry was constructed.}. An attempt to formulate a
mathematical
criterion which would select the dualizable $\si$-models was
 formulated in
\cite{OA2} where it was also illustrated how the Abelian T-duality
saturates
the criterion. We would like to stress that the  $\si$-models related by
the Poisson-Lie T-duality all fulfil the criterion of \cite{OA2}. The
modular
space of such $\si$-models is in fact huge. With arbitrary manifold
 which
has a the structure of the principal bundle with the fiber being a
Drinfeld
double we can associate dualizable $\si$-models. They are
parametrized
by an infinite space of sections of certain Grassmannian bundle
which shares
 the
same base manifold with the principal Drinfeld double
 bundle \cite{KS2}.
Because this is true for arbitrary Drinfeld
double  we see, indeed, that the modular space is
 extremely big. Only its
small corner consists of the $\si$-models admitting the
 standard non-Abelian
duals in the sense of \cite{OQ,FJ,FT}; those corresponds
 to the Drinfeld
doubles in which one of the two algebras forming the Lie
bialgebra is
commutative.

We see several reasons why to pursue the program of
investigating the
structure of the Poisson-Lie T-duality further. At the
purely classical level, we
have a huge laboratory for studying the T-duality
 without isometry and  at
the level of quantum field theory (QFT) there arises
 a possibility of existence
of non-trivial discrete symmetry relating a weak
 coupling
regime  of one QFT to  the strong coupling
regime of the dual theory. At the level of
statistical physics there is
a good chance of obtaining a natural generalization
 of the standard
Kramers-Wanier duality \cite{Itz} and last,
 but not least, a way to string
theory applications is open, since examples of the
 pairs of conformal
$\si$-models related by the Poisson-Lie T-duality
have been already
constructed \cite{AKT}.

T-duality can be interpreted as a discrete symmetry of
some conformal
field theory (CFT) whose $\si$-model description is
ambiguous. We may
say, in a sense, that the both of the dual pair
of the $\si$-models are
`present' in the CFT and they show up if we interpret
 CFT from different
(dual) points of view. The picture of CFT in terms of
 its operator
algebra and its Hilbert space we may interpret as the
duality invariant
description.  Eventually, we aim to show that the
Poisson-Lie T-duality is
a symmetry of some CFT in the same sense. A possible
 strategy for reaching
this
goal could consist first in finding   a classical
 analogue of this duality
invariant description and then in attempting a path
 integral or other
type of quantization.

 In this contribution, we shall show  that this  duality
 invariant
description
is a given in terms of a  first order (Hamiltonian) action
functional
on the phase space which is  the loop groop of the
Drinfeld double.
If we eliminate one `half' of the phase space variables
in terms
of the remaining `half'
we obtain the standard second order $\si$-model
action on the target
being  one of the (isotropic) groups forming the
 Drinfeld double.
If we choose a different description of the phase
 space and again eliminate
one `half' of the new variables in terms of the
other `half' we obtain
the dual $\si$-model living on the target of the
 dual group.
 The change of the old variables to the
new ones turns out to be nothing but a
 canonical transformation. The
generating function(al) of the canonical
transformation then follows directly
from the formalism, because it is given by a
total derivative term by
which the actions of the mutually dual models
 differ. This total derivative
term comes from the ambiguity in specifying the
 two-form whose exterior
derivative is the WZW three-form and , rather
remarkably, it is given
by the Semenov-Tian-Shansky symplectic
form on the double.

In section 2 we shall describe a detailed
form and properties of the first
order duality invariant action and show
how the dual pair of $\si$-models
can be extracted from it. In section 3 we
shall show how the formalism
directly gives the generating function(al)
of the canonical transformation
relating the models.
\section{Poisson Lie T-duality}
\subsection{`Atomic' duality}

 For the description of the Poisson-Lie
duality we need the
crucial concept
of the Drinfeld double,  which is simply
 a  Lie group $D$ such that
its Lie algebra
$\cd$, viewed as the linear space, can be
 decomposed into
a direct sum of vector spaces which are
 themselves
 maximally isotropic subalgebras with
respect to a non-degenerate invariant
bilinear form on $\cd$ \cite{D}.
 An isotropic subspace of $\cd$ is such
 that the value of the invariant form
on any
two vectors belonging to the subspace
 vanishes (maximally isotropic means
that this subspace cannot be enlarged
 while preserving its  isotropy).
 Any such decomposition of the double
 into
a pair of maximally
isotropic subalgebras
$\cg + \cgt=\cd $ is usually referred to
as the Manin triple. The Lie groups
corresponding to the Lie algebras
$\cg$ and $\cgt$
we denote as $G$ and $\ti G$.

For the sake of clarity, it is convenient first to consider  an `atomic'
Poisson-Lie T-duality which means that the targets of the $\si$-models forming
the dual pair are the groups $G$ and $\ti G$ respectively. As we have already
mentioned the phase space is the loop group of the Drinfeld double (the
elements of the loop group are loops $l(\sigma)$ in the Drinfeld double
and the multiplication is simply the point-wise one). The first order
Lagrangian is a  functional of the fields $l(\tau,\sigma)$ and it is given by
\be {\cal L}={1\over 2}\la \pas l~l^{-1},\pat l~l^{-1}\ra+{1\over 12}
d^{-1}\la dl~l^{-1},[dl~l^{-1},
dl~l^{-1}]\ra +{1\over 2}\la \pas l~l^{-1}, A\pas l~l^{-1}\ra\ee
Here $\la .,.\ra $ denotes the non-degenerate invariant bilinear form
on the Lie algebra  of the double. In the second term in the r.h.s. we
recognize
the two-form
potential
of the WZW three-form on the double and
$A$ is a linear (idempotent) map from the Lie algebra ${\cal D}$
of the double into itself. It has two eigenvalues $+1$ and $-1$, the
corresponding eigenspaces ${\cal R}_+$ and ${\cal R}_-$
have the same dimension $dimG$, they are perpendicular to each other
in the sence of the invariant form on the double and
they are given by the following recipe:
\be {\cal R}_+=Span\{t+R(t,.),t\in\cgt\},\quad  {\cal R}_-=
Span\{t-R(.,t),t\in\cgt\}.\ee
Thus the modular space of such actions is described by (non-degenerate)
bilinear forms $R(.,.)$ (matrices) on the algebra $\cgt$ \footnote{Since
$R(.,.)$ is non-degenerate, there exists
 the inverse bilinear form defined on the dual $\cg$ of
$\cgt$, hence such description of the modular space does not break the
duality.}.
For a better orientation of an interested reader we stress that the
first two terms in (1) give together the standard WZW Lagrangian on the double
if we interpret  $\tau$ and $\sigma$ as  the `light-cone' variables. These two
first terms
play the role of the `polarization'  term $pdq$ in the first order
variational principle
\be S=\int L=\int pdq-Hdt.\ee
The remaining third  term of the action (1) plays the role of the Hamiltonian
$H$. The Lagrangian (1) has also a small gauge symmetry $l\to lk(\tau)$
where $k(\tau)$ is arbitrary $\tau$-dependent function on the double.
Strictly speaking this means that the phase space is  the right coset
$LD/D$ of the loop group of $D$ by the right action of $D$ itself.
This gives some restrictions on possible zero modes of the strings.
We shall
comment the role of this fact soon.

Every element of $l\in D$ can be written as
\be l=g\ti h,\quad g\in G,\ti h\in \ti G.\ee
By inserting this   decompositions in the Lagrangian (1)
and using the Polyakov-Wiegmann formula \cite{PW}, we obtain
the following expression
$$ L=\la \Lambda, g^{-1} \partial_{\tau}g\ra
+Ad_g G(\Lambda,\Lambda) $$\be  +Ad_g G^{-1}(g^{-1}\pas g~+Ad_g
(B+\Pi(g))(\Lambda,.),
g^{-1} \pas g+Ad_g (B+\Pi(g))(\Lambda,.)).\ee
Here $\Lambda=\pas \ti h ~\ti h^{-1}$
and  we used a compact notation in order not to burden the formula with
too many indices: $G(.,.)$ and $B(.,.)$ denote respectively the symmetric
and the antisymmetric part of the bilinear form $R(.,.)$ on the Lie algebra
$\cgt$ (see Eq. (2)). $G^{-1}(.,.)$ is, in turn, the inverse bilinear form to
$G(.,.)$ and, as such, it is defined on the Lie algebra $\cg$.
 $Ad_g$ means the adjoint action of the group $G$ on the bilinear forms
and $\Pi(g)$ is a $g$-dependent antisymmetric bivector (it acts on the
elements from the algebra of $\cgt$)\footnote{ In fact, $\Pi(g)$ is nothing but
the famous Poisson bracket on $G$ which makes $G$ the Poisson-Lie group
with respect to the dual group $\ti G$.}.
$\Pi(g)$ is given by an explicit formula
\be \Pi(g) =b(g)a(g)^{-1},\ee
The matrices $a(g)$ and $b(g)$ are defined by
\be g^{-1}T^i g\equiv a(g)^i_{~l} T^l, \qquad g^{-1}\ti T_j g \equiv
b(g)_{jl}T^l +d(g)_j^{~l} \ti T_l,\ee
in (7) the adjoint action of $G$ is understood to take place in the Drinfeld
double algebra ${\cal D}$ and the pair of  bases
$T^i$ and $\ti T_i$ in the algebras $\cg$ and
$\cgt$ respectively, satisfy the duality condition
\be \langle T^i,\ti T_j\rangle =\delta^i_j.\ee

Note that $\Lambda$ appears quadratically in the Lagrangian (5). One might
think that it could be directly integrated from the Lagrangian even in
the path integral sense but the story is more complicated , however. The reason
is that our strings $l(\tau,\sigma)$ in the double are closed. This fact gives
the following constraint  of the unit monodromy on $\Lambda$:
\be P\exp{\int_{\gamma} \Lambda}=\ti e,\ee
where $P$ stands for the path-ordered exponential, $\gamma$ is a closed
path around the string worldsheet and $\ti e$ is the unit element of the
dual group.
 Thus, though the path integral with the Lagrangian (5)
seems to be Gaussian,  the non-local constraint (9) makes it
more difficult to compute. In fact, we have not found a way how to compute
it yet though we believe it can be eventually managed in future. Here we just
remark that we can proceed at the level of equations of motions. Indeed, let us
vary $\Lambda$ while preserving the constraint (9) . We  obtain $\Lambda$ as a
function of $g$. Inserting this function $\Lambda(g)$ back in (5) we get
a $\sigma$-model Lagrangian on the target $G$:
\be L=(R+\Pi(g))^{-1}(\partial_+ g g^{-1}, \partial_- g g^{-1}),\ee
where
\be  R(.,.)=G(.,.)+B(.,.), \qquad
\partial_{\pm}=\pat\pm\pas.\ee
We should keep in mind, however, that the solution of the equations
of motion coming from the Lagrangian (10) should be subject to the constraint
(9) expressed in terms of $\Lambda(g)$. It is precisely this constraint
which  from the point of view of the $\sigma$-model (10) on $G$ entails
the factorization $LD/D$ mentioned before.
 Note that the model (10) possesses
an isometry with respect to the right action on the group manifold only
in case when the dual group is commutative (the Poisson bracket $\Pi(g)$ then
vanishes).

Now we choose the `dual' parametrization of an element $l$ of the double:
\be l=\ti g h, \quad \ti g\in \ti G, h\in G.\ee
Whole procedure of integrating $h$ away can be repeated without any
change  and we arrive at the obviously equivalent dual $\sigma$-model
\be \ti L=(R^{-1}+\ti \Pi(\ti g))^{-1}(\partial_+ \ti g \ti g^{-1},
\partial_- \ti g \ti g^{-1}).\ee
The pair of the mutually dual $\sigma$-models (10) and (13) was constructed
in \cite{KS2} by using a geometric picture of lifting the extremal
strings of the model (10) to the double and then projecting them on the
extremal configurations of the dual model (13). The derivation presented
in the present note clearly shows the common `roof' (1) of the two models.
So we may conclude that the Poisson-Lie T-duality just exchange the
two group manifolds and the matrix
$R$ by its inverse. In case the both groups dual to each other are the same
(Abelian case and also the case of the so-called Borelian Drinfeld doubles) the
duality amounts just in exchanging the matrix $R$ by the matrix $R^{-1}$.
We shall give some concrete examples of the  dual pairs of model after
the discussion of the canonical transformations in the general case.

Note that for every decomposition of the Lie algebra of the
double ${\cal D}={\cal K}+\ti{\cal K}$ into two subalgebras with the
property $\la {\cal K},{\cal K}\ra=\la \ti{\cal K},\ti {\cal K}\ra =0$,
we again generate a pair of the dual $\sigma$-models on the corresponding
group targets $K$ and $\ti K$. Thus the action (1) decribes a whole space
of the equivalent $\sigma$-models; the space is simply given by the set
of all such isotropic decompositions of the algebra of the double. In the
case of the Abelian duality it is given by the orbit of the
group $O(d,d;Z)$. In the next section we shall demonstrate how to use
our results to find the generating functions of the canonical transformations
among the equivalent models.
\subsection{Buscher's duality}
In this paragraph we give a very brief description of Buscher's duality
with the purpose just to illustrate to an interested reader that the
Poisson-Lie
duality is not just duality between two group targets but it applies in much
more
general setting. The corresponding formalism is more cumbersome
than in the case of the atomic duality and we plan to give its most detailed
account
in a separate publication. Here we sketch just the results.
Consider some manifold $M$ and some coordinates $x^{\mu}$ on it. Consider
then its cotangent bundle $T^*M$ and  the coordinates $x^{\mu},p^{\mu}$ which
are the Darboux coordinates in which the canonical symplectic form $\omega$ on
$T^*M$
looks like $\omega=dp_{\mu}\w dx^{\mu}$.  We may construct the following
first order Lagrangian on a phase space formed by closed loops in the  manifold
$T^*M\times D$:
\be  {\cal L}={1\over 2}\la \pas l~l^{-1},\pat l~l^{-1}\ra+{1\over 2}
d^{-1}\la dl~l^{-1}[dl~l^{-1},
dl~l^{-1}]\ra +p_{\mu}\pat x^{\mu} +{1\over 2}\la J, A(x)J\ra.\ee

Here the bracket  in the last term in the r.h.s. $\la .,.\ra $ denotes a
non-degenerate bilinear form
on the  tangent spaces to $T^*M\times D$
at points  living on the submanifold
$M\times D$ \footnote{If  $S,S'$ are some two elements from
such a tangent space at some point $x$ of the submanifold $M\times D$ ;
for instance  $S=(t,\beta,\rho); t\in T_x M, \beta\in T_x^*M,\rho\in{\cal D}$
and   $S'=(t',\beta',\rho'); t'\in T_x M, \beta'\in T_x^*M,\rho'\in{\cal D}$
then
$$ \la S,S'\ra = \la \beta,t'\ra+\la \beta',t\ra +\la \rho,\rho'\ra.$$
Here the first two brackets on the r.h.s. mean the standard pairing between
$T_x^*M$
and $T_x M$ and the third bracket is the standard bilinear form (8) on the
double.}.
$A(x)$ is a linear (idempotent) map from the space  $T_x^*M +T_x M+ {\cal D}$
 into itself. It has two eigenvalues $+1$ and $-1$, the
corresponding eigenspaces ${\cal R}_+(x)$ and ${\cal R}_-(x)$
have the same dimension $dimG+dimM$, they are perpendicular to each other
in the sence of the just described bilinear form on   $T_x^*M +T_x M+ {\cal D}$
and
they are given by the following recipe:
\bea{\cal R}_+(x)&=Span\{t+R(x)(t,.),t\in T_x ^*M+\cgt\},\cr  {\cal R}_-(x)&=
Span\{t-R(x)(.,t),t\in T_x^*M +\cgt\}.\eea
Thus the modular space of such actions is described by (non-degenerate)
bilinear forms $R(x)(.,.)$ (matrices) on the space  $T_x ^*M+ \cgt$
\footnote{Since  $R(x)(.,.)$ is non-degenerate, there exists
 the inverse bilinear form defined on the dual  space $T_x M +\cg$ of $T_x^*M+
\cgt$, hence such description of the modular space does not break the duality.}
or, in other words, by sections of a Grassmannian bundle over the base manifold
$M$. Finally, it
remains to explain what is $J$ in (14).Of course, it is an element  of
$T_{x(\si)}M+T_{x(\si)}^*M+{\cal D}$
given by
\be J=(\pas x^{\mu}(\si), p_{\mu}(\si),\pas ll^{-1}),\ee
where $x^{\mu}(\si),p_{\mu}(\si),l(\si)$ is  an element of the phase space of
the string.

As in the case of the atomic duality, we first write $l=g\ti h$ and  solve away
from the
action the fields $p_{\mu}$ and $\ti h$. We obtain the following  $\si$-model
Lagrangian
\be L=(R(x)+\Pi(g))^{-1}(j_+, j_-),\ee
where  (with a slight abuse of the notation) $\Pi(g)$ now denotes a bilinear
form
on the space  $T_x ^*M+\cgt$ which on $\cgt$ acts as  $\Pi(g)$ in (6)  before
and on $T_x^*M$ it vanishes;
$j_{\pm}$ are elements of  $T_x M+\cg$ given by
\be j_{\pm}=(\partial_{\pm} x^{\mu},\partial_{\pm}g g^{-1}).\ee
We may also write $l=\ti g h$ and solve away the fields $p_{\mu}$ and $h$. The
result is the
Lagrangian of the dual $\si$-model
\be \ti L=(\ti R(x)+\ti \Pi(\ti g))^{-1}(\ti j_+, \ti  j_-),\ee
where
\be \ti j_{\pm}=(\partial_{\pm}x^{\mu},\partial_{\pm}\ti g \ti g^{-1})\ee
and $\ti R(x)$ is given in terms of $R(x)$ as
\be \ti R(x)=(A+R(x)B)^{-1}(C+R(x)D).\ee
Here ($Id$ means the identity matrix)
\be A=D=\left(\matrix{Id&0\cr0&0}\right),\qquad B=C=
\left(\matrix{0&0\cr0&Id}\right)\ee
and the block matrices are understood in terms of the decomposition $T_x^*M
+\cgt$.
Needless to say, in the standard Abelian case ($D=U(1)\times U(1)$)  the
formula (21)
gives just the  Buscher transformations \cite{Busch}.

\section{Canonical transformations}
As we have already mentioned we can interpret the duality invariant first
order action as the sum of the WZW action (which corresponds to the
polarization form $pdq$ on the phase space of the model) and the expression
quadratic in the Drinfeld double
currents  (which plays the role of the Hamiltonian). We can  parametrize
the
first order action in terms of the fields $g,\ti h$ and $\ti g, h$
respectively such that $g\ti h=\ti g h =l$ and we obtain, after eliminating
$\ti h$ or $h$, the dual pair of the $\si$-models  (10) and (13). The
transformation
between those two parametrizations is obviously canonical transformation
because it respects the form $pdq-Hdt$ of the variational principle, moreover,
it preserves the Hamiltonian itself. Hence, by definition, the total derivative
of the
generating function is given by the difference $PdQ-pdq$ where the capital
characters denote the new coordinates ($\ti g(\si),h(\si)$)  and the small
characters the old
coordinates ($g(\si),\ti h(\si)$).  In our case we have used the
Polyakov-Wiegmann
formula\footnote{Note that the WZW action $I$ of a {\it single} argument
$g,\ti g,h$ and $\ti h$ vanishes.}
\be I(\ti gh)=\int \la \pas h h^{-1}, \ti g^{-1}\pat \ti g \ra  d\tau d\sigma
;\ee
\be I(g\ti h)=\int \la \pas \ti h \ti h^{-1}, g^{-1}\pat g \ra  d\tau d\sigma
\ee
to obtain the explicit form of the terms in the actions  corresponding
to polarization forms $PdQ$ and $pdq$. We know already
by construction that  the two expressions in the r.h.s. of (23) and (24) at
most differ
by a total derivative (they may not be identical because of the ambiguity in
defining $d^{-1}$ of the WZW three-form; this ambiguity is a total derivative).
In principle, it is now easy task to evalute the generating function(al)
$F(g,\ti g)$ of the canonical
transformation between  $\ti g, h$ and $g, \ti h$. It is given by
\be  \int  dF =\int  \la \pas h h^{-1}, \ti g^{-1}\pat \ti g \ra  -
\int \la \pas \ti h \ti h^{-1}, g^{-1}\pat g \ra, \quad h=h(g,\ti g), \ti h
=\ti h(g,\ti g).  \ee
The dependence of $h,\ti h$ on $g,\ti g$ is, of course, given by the
requirement
encountered before: $g\ti h=\ti g h$. Thus we have obtained  an implicit
expression for the
generating function(al)  $F$;  to make it explicit for a concrete  Drinfeld
double we
only have to express $g$ and $\ti g$ in terms of $h$ and $\ti h$.   Remarkably,
however,  it is possible to write a  more explicit formula for  $F$  still
considering a generic Drinfeld double.
It was derived already  in \cite{KS2}  (that time we did not know yet  the
duality
invariant action (1) on the double). The result is
\be F(g,\ti g)=\int _{D(l)} \Omega.\ee
Here  $g(\si),\ti g(\si)$ parametrize the elements $l(\si)$ of the phase space
$LD$ of the model
in such a way that   $h(g,\ti g)$ and $\ti h(g,\ti g)$ interpolate between $g$
and $\ti g$,
i.e. $l=g\ti h=\ti g h$;  $D(l)$  is an arbitrary two-dimensional surface
embedded in the
double whose boundary is just the loop $l(\si)$  and  $\Omega$ is the
symplectic (hence closed)
two-form on the double constructed by Semenov-Tian-Shansky\footnote {The form
$\Omega$ has the important property that the action of the double with the
standard {\it Poisson}
structure on the double with the Semenov-Tian-Shansky  {\it symplectic}
structure is the Poisson
action.}\cite{SM}. It is easy to describe the form $\Omega$ in terms of the
Poisson bracket on the double
to which it gives rise.  Define $(\nabla_L f)_a,(\nabla_L f)^a,(\nabla_R f)_a$
and
$ (\nabla_R f)^a$
as follows
\be df =(\nabla_L f)_a (dl l^{-1})^a +(\nabla_L f)^a (dl l^{-1})_a=(\nabla_R
f)_a (l^{-1} dl)^a +(\nabla_R f)^a ( l^{-1} dl )_a,\ee
where $f$ is some function on the double. Clearly, the upper and lower indices
for the forms $dl l^{-1}$ (or  $l^{-1} dl$) mean:
\be dl l^{-1} =(dl l^{-1})_a T^a +(dl l^{-1})^a \ti T_a.\ee
Then the Semenov-Tian-Shansky Poisson bracket is given by
\be \{f,f'\}=\Omega^{-1}(df,df')=(\nabla_L f)_a(\nabla_L f')^a-(\nabla_R
f)^a(\nabla_R f')_a\ee
for arbitrary functions $f,f'$ on the double.

The formula (26) gives the maximally explicit description of the generating
function
of the canonical transformation which can be obtained for a {\it generic}
double.
It uses only the canonical structure on a Drinfeld double such as the
Semenov-Tian-Shansky
form certainly is. In concrete examples of the Drinfeld doubles  with concrete
parametrizations
of their  group manifolds the formula (26) for the generating function  look in
general very cumbersome.
We feel, however, that it  is its rather positive than negative feature of the
formula (26). Indeed,
we have some doubts that it would be easy to find the formula for the
generating
function without having the geometric understanding ot  $T$-duality which has
led to the
conceptually simple formula (26).
It is also worth mentioning  that the dual $\si$-models (in the sense of the
Buscher duality)
 (17) and (19) are also connected by a canonical transformation whith the same
generating
function $F(g,\ti g)$ given by (26). In other words: the structure
of the `attached' manifold $M$ with the coordinates $x^{\mu}$ is highly
irrelevant for
the duality transformation.

The duality invariant first order action (1) enables us to find the formula for
the
generating function of  the canonical transformation between arbitrary pair of
mutually
equivalent $\si$-models coming from the action (1) for all possible choices
$({\cal K},\ti {\cal  K})$ of the isotropic decompositions of the Lie algebra
of the double
(see the discussion in the previous section).  Consider e.g.  some two
decompositions: ${\cal D}=\cg +\cgt$ and ${\cal D}={\cal K}+\ti {\cal K}$ and
find the
generating function of
the canonical transformation between the $\si$-models, say, on target $G$ and
target $K$.
In order to do that  we parametrize $l\in D$ as $l=g\ti h=k \ti m$ where $g\in
G$,$\ti h\in \ti G$,
$k\in K$ and $\ti m\in \ti K$. The generating function $F(g,k)$ is then  given
by
\be  \int  dF =\int  \la \pas \ti m \ti m^{-1}, k^{-1}\pat  k \ra
-\int \la \pas \ti h \ti h^{-1}, g^{-1}\pat g \ra, \quad \ti m=\ti m(g,k), \ti
h =\ti h(g,k).  \ee

\vskip 1pc

\noindent {\bf Acknowledgement.} We thank A. Alekseev, L. \'Alvarez-Gaum\'e,
E. Kiritsis, R. Stora. A. Tseytlin and E. Verlinde  for discussions.
\vskip 1pc
\noindent{\bf Note added:} In the few hours' gap between completing the present
paper
and submitting it to hep-th we received a preprint \cite{TU}  where an
 interesting and
differently looking
form of  a  Poisson-Lie duality invariant action was also presented. We hope to
elucidate the relation between those two actions in near future.
We believe, however,  that another
 main result of \cite{TU} which should be the path integral derivation
of the Poisson-Lie T-duality  requires more rigour. The reason is that the
authors
of \cite {TU} did not  comment that the fields  over which they integrate are
also constrained
by the highly non-linear and non-local unit monodromy constraint (9) and
therefore
the process of continual integration over such fields should become more
non-trivial.
This is also the reason why we hesitate to claim that we already have  a path
integral
derivation of the Poisson-Lie T-duality.

\end{document}